\documentclass{article}
\usepackage[english]{babel}
\usepackage[letterpaper,top=2cm,bottom=2cm,left=3cm,right=3cm,marginparwidth=1.75cm]{geometry}
\usepackage{amsmath,amssymb}
\usepackage{graphicx}
\usepackage[colorlinks=true, allcolors=cyan]{hyperref}
\usepackage{lipsum} 
\usepackage{authblk}
\usepackage{url}
\usepackage{hyperref}
\usepackage{color}
\usepackage[normalem]{ulem} 
\usepackage{pbox} 
\usepackage{makecell}
\usepackage{pbox}
\usepackage{tabulary}
\usepackage{array}
\usepackage{rotating}
\usepackage{tikz}
\usepackage{colortbl} 
\usepackage{xcolor}
\usepackage{multicol}
\usepackage{caption}
\usepackage{subcaption}
\usepackage{tikz}
\usepackage{amsmath}
\usepackage{graphicx}
\usepackage{caption}
\usepackage{pgfmath}

\providecommand{\keywords}[1]
{
   	
  \textbf{{Keywords:}} #1
}

\title{Viral mutation spread controlled by inter-host network dynamics}

\author[1,2,3]{\underline{Javier L\'opez-Pedrares}}
\author[1,3]{M. Elena V\'azquez-Cend\'on}
\author[1,2,\thanks{corresponding author, email: \href{mailto:alberto.perez.munuzuri@usc.es}{alberto.perez.munuzuri@usc.es} }]{Alberto P. Mu\~nuzuri}
\affil[1]{Galician Center for Mathematical Research and Technology (CITMAga), 15782 Santiago de Compostela, Spain}
\affil[2]{Group of Nonlinear Physics, Universidade de Santiago de Compostela, 15782 Santiago de Compostela, Spain}
\affil[3]{Department of Applied Mathematics, Universidade de Santiago de Compostela, 15782 Santiago de Compostela, Spain}

\begin{document}
\date{}
\maketitle
\begin{abstract}

The increase in the connectivity between hosts in recent times  
has facilitated the emergence of more aggressive mutant viral strains, making their containment and eradication significantly more challenging compared to the original variants. We focus on the evolution of a new more aggressive mutant strain that appears in the system and competes with a previous version. 
The role of host interaction network topology in the emergence and spatial diffusion of these highly contagious strains is analyzed, as well as how network-based interventions can help mitigate their spread. To address these issues, we present simplified mathematical models that qualitatively describe the occurrence, propagation, and impact of such mutations within host-interaction networks. By incorporating the topology of the host network into the analysis, a more advanced framework is proposed to curb the growth of mutant strains. Extensive numerical simulations of these models offer insights into the mechanisms driving viral evolution, the dynamics of contagion, and the topological influence of host networks on viral spread and containment strategies.

\end{abstract}


\section{Introduction}
\label{sec:intro}

The spatial dynamics of infectious diseases have been studied in depth in recent decades \cite{pastor2001epidemic,min2018competing,van2013contagious}. The scenarios in which these diseases develop have changed drastically  \cite{strogatz2001exploring} and different mathematical models have emerged that try to mimic this complex phenomenon. Furthermore, with the emergence of mutations within the same host, the complexity of the problem has increased. 

The competitive exclusion principle  \cite{hardin1960competitive} states that two species cannot coexist if they compete for identical resources. In the context of a viral infection, two strains of the virus—one original and other mutant—compete for the same resources within the host.  This competition will likely result in the survival of the more fit strain, while the less suited strain will be out-competed and eventually driven to extinction.

In this letter, we examine the competition between two virus strains, the original and a new mutant more-aggressive one, and the influence of host-to-host network structures on the evolutionary dynamics of this competition. Building upon the framework of the potential host network introduced in \cite{lopez2023interactions}, this study extends the analysis to explore how these networks shape viral evolution. Notably, viruses often undergo mutations to adapt to their environments, periodically resulting in the emergence of more virulent strains. The findings presented here underscore the critical role of network topology in controlling the spread of a new more-aggressive virus variant along the hosts aiming to control its propagation and the time to reach dominance.

The document is organized as follows. In the methodology section we present the coupled model describing the mutant-variants competition within a host coupled with the model describing the spread of the disease among the potential hosts. This section also contains information on the different networks describing the interaction between potential hosts. The numerical solution of the problem is presented in the results. The final section discusses and draws conclusions.

\section{Methodology}
\label{sec:methodology}

In order to introduce the effect of host-to-host network topology in disease models of infection and competition we build on the model already studied in \cite{lopez2023interactions}. The proposed mathematical approach mimics the scenario in which two lineages of the same virus cohabit in an organism and spread through a network of hosts.

The formulation and construction of the coupled model are comprehensively detailed herein. The theoretical framework, along with the integration of its constituent subsystems, is systematically described to ensure an accurate representation of the coupled phenomena. Particular attention is given to the mechanisms of interaction between the subsystems and the methodologies employed to achieve a consistent coupling.

\subsection{Full coupled model}
\label{subsec:full_model}

The epidemiological model considered in the study is based on the classical SIR compartmental model \cite{kermack1927contribution,diekmann2000mathematical,hethcote1989three}. The hosts  considered can express any of the following three states, depending on the stage of infection: susceptible ($S$), infected ($I$) and recovered individuals ($R$). In our study, we  take into account that after a latency period, those recovered can be re-infected. These models have been previously studied in the literature \cite{ndairou2020mathematical,gomes2004infection,katriel2010epidemics} and are known as reinfection models.

The mutant-variants model used is based on the classic Lotka-Volterra competition model \cite{wangersky1978lotka}. A model of competition between the original lineage and a more-aggressive mutated one, can be described by the following set of ordinary differential equations \cite{fabre2012modelling},

\begin{equation}
\begin{split}
    \frac{dV_A(t)}{dt} = r_A \left( 1 - \frac{V_A(t) + V_B(t)}{k_A} \right) V_A(t), \quad
    \frac{dV_B(t)}{dt} = r_B \left( 1 - \frac{V_A(t) + V_B(t)}{k_B} \right) V_B(t),
\end{split}
\label{original}
\end{equation}

\noindent
where $r_A, r_B, k_A, k_B \in \mathbb{R}^+$ and $V_i(t)$ denotes the viral density of species $i$ at time $t$, $r_i$ denotes the reproduction rate of virus $i$, and $k_i$ denotes the competition rate of virus $i$, with $i = A, B$. We consider $k_A < k_B$ to ensure that the second strain is the  mutant more-aggressive strain. 

Throughout this manuscript we propose to study the effect of interactions between virus-carrying hosts and how the topology of these interactions can be used to control the evolution of both viruses. The interactions described above constitute a complex network whose nodes are the hosts and the connections represent the relationships between two individuals that can transmit the infections. Mathematically, it is easy to describe this network by means of the adjacency matrix, $\mathcal{A}$ ($\mathcal{A}_{ij} = 1$ if individual $i$ interacts with $j$ and $0$ otherwise). Note that the adjacency matrix is symmetric by construction, moreover in our problem we consider non-weighted networks \cite{barrat2008dynamical}. The three network topologies examined are as follows: small-world (Watts-Strogatz), scale-free (Barabási-Albert), and random networks (Erdös-Rényi), each representing a distinct form of interaction among organisms \cite{estrada2012structure,mata2020complex,van2010graph}.

All the components described above have been integrated into a single model. The resulting model is an integrated model that encompasses the idea of contact model \cite{bansal2010dynamic}. The host-to-host network encloses the connections between individuals who may become infected. Such contagions are given by the SIR model with reinfection. Finally, the mutant-variants competition model marks the temporal evolution of the two viruses infecting the population.

The network is initialized by assigning a subset of nodes as infected (typically $0.3\%$ of the nodes), with the remaining nodes (of a total of $1000$) in the susceptible state. The original strain $k_A$ is always kept at $k_A=10$ throughout the manuscript. The simulation proceeds iteratively, at each time step, the probability of an infected individual transmitting the disease to susceptible neighbors is evaluated using a Monte Carlo method. Similarly, the recovery probability for each infected individual is computed at the end of the time step. The mutant competition dynamics, governed by \eqref{original}, is numerically integrated at each step using an implicit Euler method. This process continues until the infected population is eradicated or a unique virus dominates.

\section{Results}
\label{results}

We performed a large, statistically significant, number of simulations of the model presented above. In all simulations, the networks considered have one thousand nodes. In our case study, we only have two strains, original and mutant, so the computation times are negligible (1 minute) compared to simulations with a large number of strains, as shown in \cite{lopez2023interactions}. During simulations, a key metric is recorded: the dominance time  of a single virus, $t_{\text{fin}}$. This parameter measures the time it takes for the system to verify the Competitive Exclusion Principle, thus, only one virus remains active and the program is finalized. We also track another relevant parameter that we denote by $t_{\text{dom}}$ or time for the dominant (mutated) strain to be present in all the infected hosts. At $t_{\text{dom}}$ traces of the weaker strain might still be present in the system, thus, its value is always lower or equal to $t_{\text{fin}}$ ($t_{\text{dom}} \le t_{\text{fin}}$).

\begin{figure}[b!]
    \centering
    \includegraphics[width=0.70\textwidth]{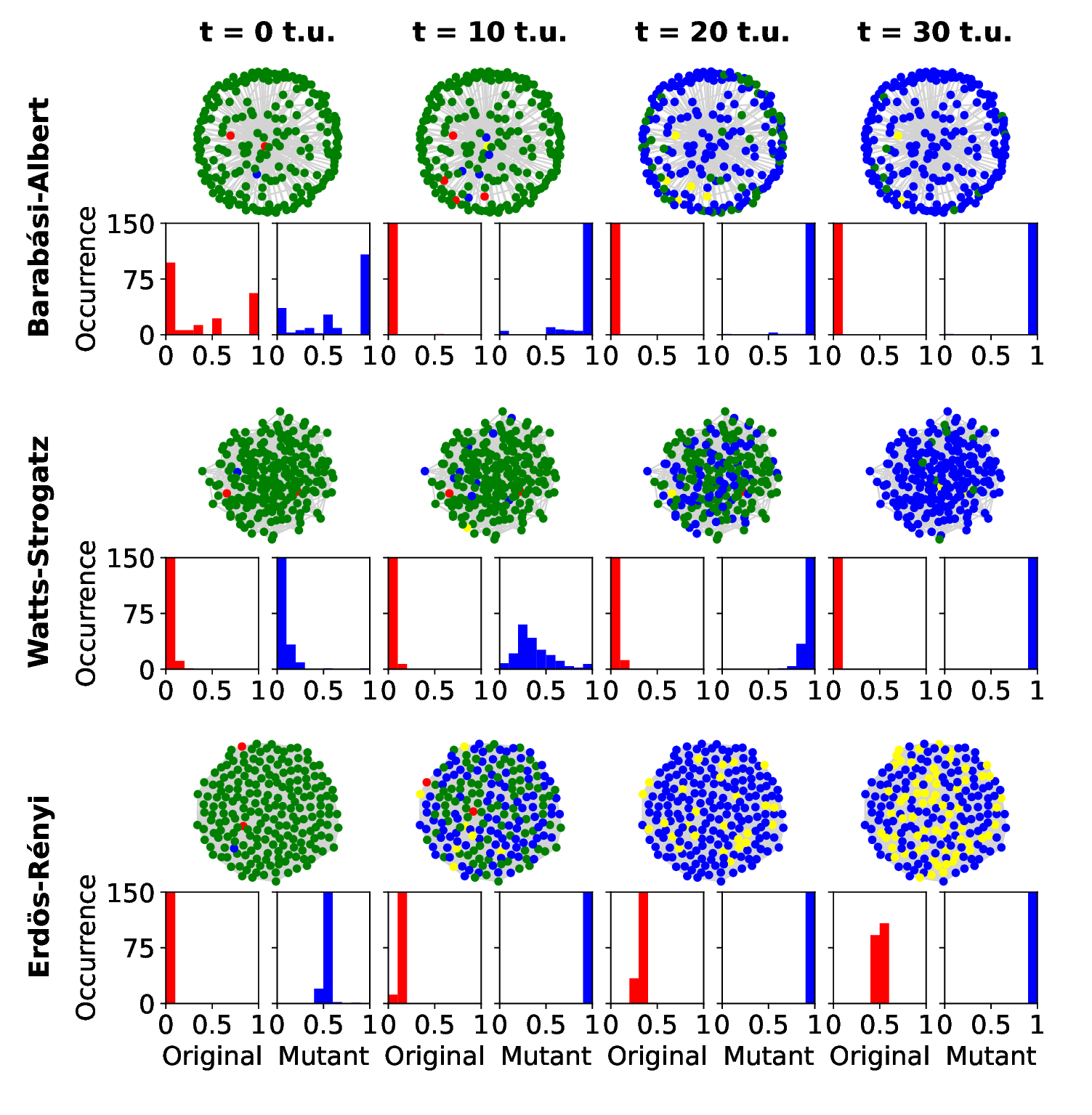}
    \caption{Evolution of the epidemic for the three networks considered. Nodes in green correspond with non infected hosts, in red with hosts infected with the original strain, in blue with the mutant strain and in yellow with both strains. Only one fifth of the total nodes are plotted to simplify the visualization. The simulations corresponds to $k_B=50$.}
    \label{fig:evolution}
\end{figure}

As introduced in Section \ref{sec:methodology}, three different types of networks have been used in the simulations, corresponding to three completely different ways of interaction between hosts. Examples with these networks are plotted in Figure~\ref{fig:evolution}. The upper row shows the temporal evolution of the spatial distribution of infected hosts for a Barabási-Albert network, the third row are the equivalent results for a Watts-Strogatz network (with $p=0.9$) while the fifth row shows the results for an Erdös-Rényi network. Nodes in green correspond
with non infected hosts, in red with hosts infected with the original strain, in blue with the mutant strain and in yellow with both strains. 
Right bellow each network, we plot histograms with the viral density for the original and mutated strains. 
A first look at these figures demonstrates that the evolution of viruses strongly depends on the network topology, as demonstrated elsewhere \cite{lopez2023interactions}. Erdös-Rényi topology helps the infection to spread throughout the system while the Watts-Strogatz topology helps containing the spread. It is the analysis of the histograms in the second row for each type of network that provides information about the competition between the two strains. In all cases, at $t=30 t.u.$ the mutated virus dominates and the other almost goes extinct. Nevertheless, the process to reach this point strongly depends on the network topology. For the Erdös-Rényi network, the original strain remains active till the end of the simulations, thus containing its spread, while for the two other topologies it becomes practically extinct at $t=20 t.u.$

\begin{figure}[t!]
    \centering
    \includegraphics[width=0.9\textwidth]{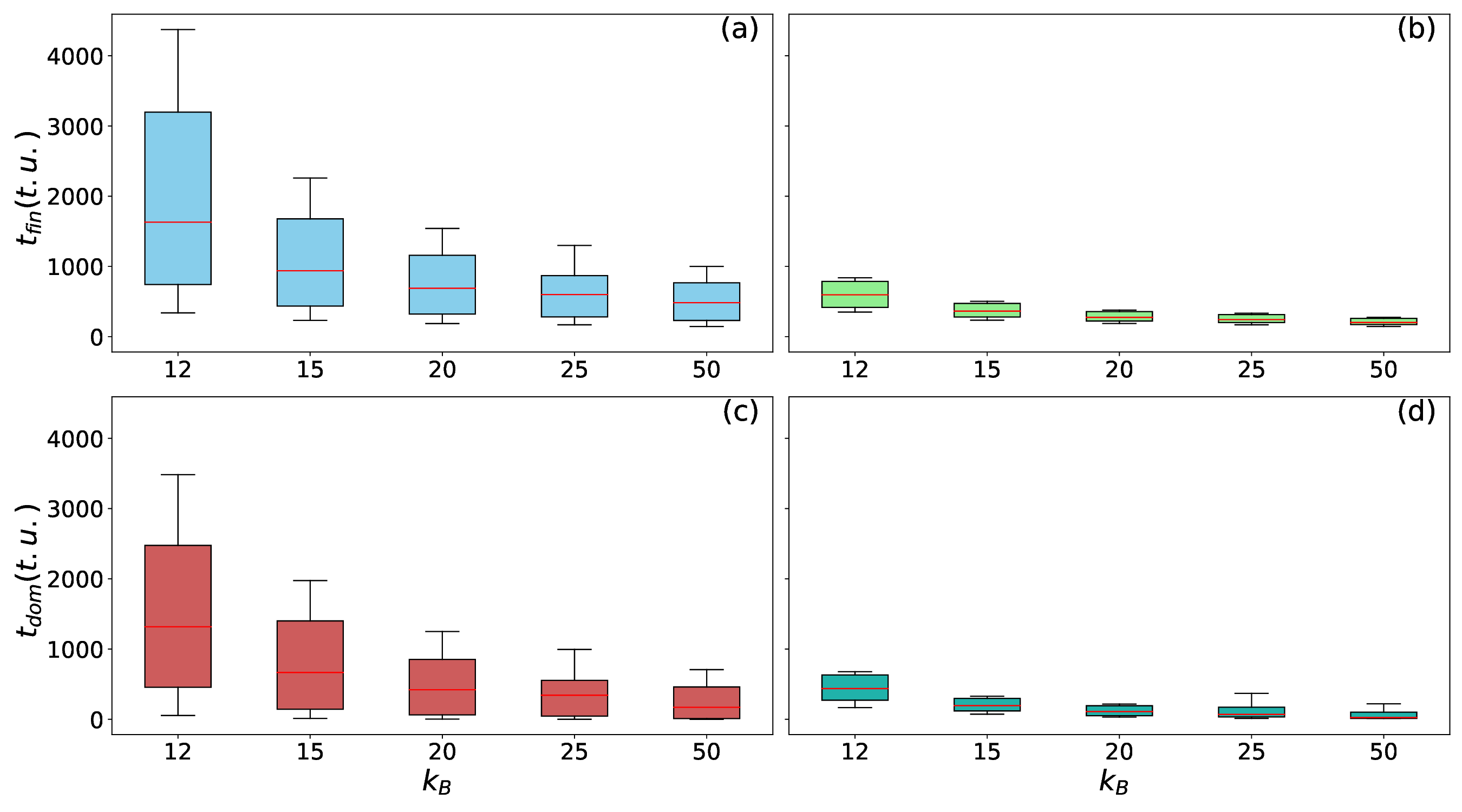}
    \caption{Comparison for different networks of box plots showing the median of the dominance time for different mutation values. Values for (a) and (c) Barabási-Albert networks and (b) and (d) Erdös-Rényi networks.}
    \label{fig:bara_rand_boxplot_vf}
\end{figure}

In Figure~\ref{fig:bara_rand_boxplot_vf} a comparison is made between two of the above networks: Barabási-Albert (Fig~\ref{fig:bara_rand_boxplot_vf}a and Fig~\ref{fig:bara_rand_boxplot_vf}c) and Erdös-Rényi (Fig~\ref{fig:bara_rand_boxplot_vf}b and Fig~\ref{fig:bara_rand_boxplot_vf}d) networks considering different degrees of mutant virus aggressiveness (given by $k_B$). 
We show box plots representing the distribution of the times, $t_{\text{fin}}$ (upper row), and $t_{\text{dom}}$. For Barabási-Albert networks both $t_{\text{dom}}$ and $t_{\text{fin}}$ are clearly longer than for the Erdös-Rényi network. A fully random network lacks to control the spread of a epidemics as expected. We can also observe that the values of $t_{\text{dom}}$ follow the same trend as those for $t_{\text{fin}}$. It means that a good way to control the introduction of the new mutant strain or, at least, delay it, is to make the hosts to interact via a Barabási-Albert network.

\begin{figure}[t!]
    \centering
    \includegraphics[width=0.9\textwidth]{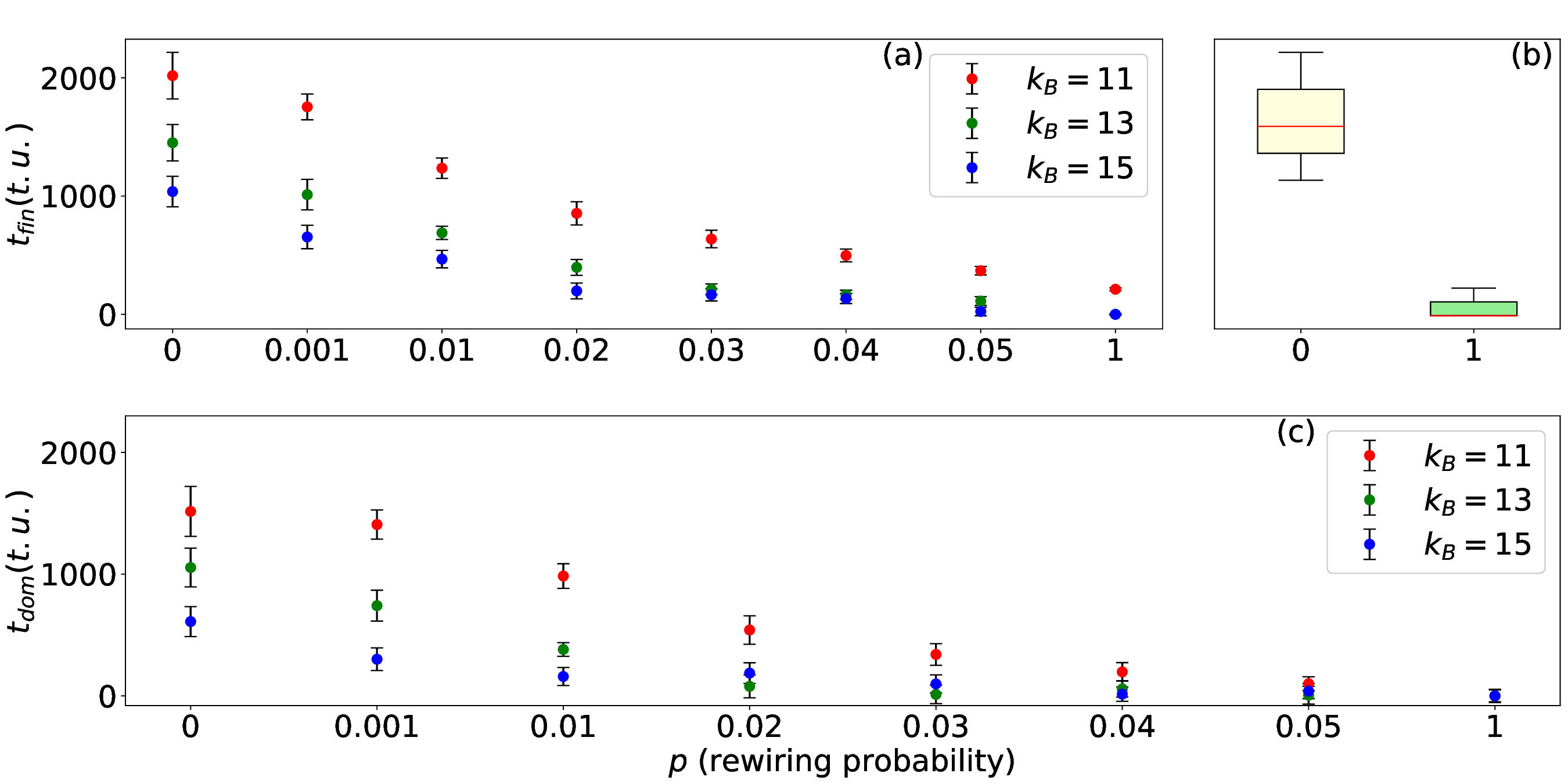}
    \caption{(a) Mean and standard deviations for $t_\textbf{fin}$ and (c) for $t_\textbf{dom}$ for Watts-Strogatz type networks considering different aggressiveness of the mutant lineage. (b) Box plots showing the median of the dominance time for different rewiring probabilities with $k_B=20$.}
    \label{fig:wattsvf}
\end{figure}

Figure~\ref{fig:wattsvf} shows equivalent results for a Watts-Strogatz network considering different values for the rewiring parameter $p$ that describes the percentage of links between non-adjacent nodes. $p$ can describe all types of networks from a regular one up to a completely random one.
The upper row in Figure~\ref{fig:wattsvf} shows the values of $t_{\text{fin}}$ versus $p$ for three different values of the mutant aggressiveness $k_B$. The lower row shows the equivalent results for $t_{\text{dom}}$ or time at which the mutant strain is present in any infection. It also seems clear from this figure that the mutant strain delays its presence by reducing $p$.

Note that in all simulations and independently of the type of interaction between potential hosts, the surviving virus is always the same. Mutant strains (by election) with larger value of $k$ are more prone to dominate than the original lineage.  In addition, the initial choice of infected hosts has been found to be indifferent. This is also independent of the type of network considered, due to the aggressiveness of the mutations.

\section{Conclusions}

Along the manuscript, a model has been presented that describes how a mutant more aggressive strain spreads through a host network competing at the same time with a previously existing less aggressive strain. We focused on the spatial distribution of the infections on the hosts and how the topology of the network controls the spread and evolution of the viral strains as well as the effective presence of the mutated potentially-more-dangerous strain in the epidemics.

As expected, the model considered recovers one of the fundamental principles of Biology: the Principle of Competitive Exclusion, so that a single strain survives when competing for the same resources. In our simulations, the mutant strain, the one with the higher $k$ rate, always survives.

The effect of the interaction network on the model is clear. The time it takes to reach the stationary state, i.e. the dominance of a single viral strain, depends on the topology of the network, where the most connected networks are the ones that reach this state first. Note that $t_{\text{dom}}$ follows the same trend, meaning that the introduction of the new more-aggressive mutated strain is also controlled by the topology of the hosts interactions. This opens the possibility to control the spread of a more-aggressive mutated strain by means of controlling the topology of interactions between the hosts.

On the other hand, the effect of the aggressiveness of the mutation was checked. Increasing the aggressiveness of the mutant stain (by increasing the value of $k$) shortens the time to dominate.

In conclusion, the network topology or spatial distribution of individuals and their interactions drastically changes the behavior of an epidemic allowing to control the temporal evolution of new mutated viral strain potentially helping to delay its spread through the population and helping to gain time to develop specific treatments.

\section*{Declaration of competing interest}
The authors declare that they have no known competing financial interests or personal relationships that could have appeared to influence the work reported in this paper.

% \section*{Data availability}
% No data was used for the research described in the article. 

\section*{Acknowledgments}
APM and JLP gratefully acknowledge financial support by the Spanish Ministerio de Economía y Competitividad and European Regional Development Fund under contract PID 2020-113881RB-I00 AEI/FEDER, UE, and by Xunta de Galicia under Research Grant No. 2021-PG036. All these programs are co-funded by FEDER (UE). EVC received financial support from the Xunta de Galicia (2021 GRC GI-1563 - ED431C 2021/15)
and this work has been partially supported by FEDER, Ministerio de Ciencia e Innovación-AEI research project PID2021-122625OB-I00.

\bibliographystyle{unsrt}
\bibliography{sample}

\begin{thebibliography}{10}

\bibitem{pastor2001epidemic}
Romualdo Pastor-Satorras and Alessandro Vespignani.
\newblock Epidemic spreading in scale-free networks.
\newblock {\em Physical review letters}, 86(14):3200, 2001.

\bibitem{min2018competing}
Byungjoon Min and Maxi San~Miguel.
\newblock Competing contagion processes: Complex contagion triggered by simple
  contagion.
\newblock {\em Scientific reports}, 8(1):10422, 2018.

\bibitem{van2013contagious}
Willem~G Van~Panhuis, John Grefenstette, Su~Yon Jung, Nian~Shong Chok, Anne
  Cross, Heather Eng, Bruce~Y Lee, Vladimir Zadorozhny, Shawn Brown, Derek
  Cummings, et~al.
\newblock Contagious diseases in the united states from 1888 to the present,
  2013.

\bibitem{strogatz2001exploring}
Steven~H Strogatz.
\newblock Exploring complex networks.
\newblock {\em nature}, 410(6825):268--276, 2001.

\bibitem{hardin1960competitive}
Garrett Hardin.
\newblock The competitive exclusion principle: an idea that took a century to
  be born has implications in ecology, economics, and genetics.
\newblock {\em science}, 131(3409):1292--1297, 1960.

\bibitem{lopez2023interactions}
Javier L{\'o}pez-Pedrares, M~Elena V{\'a}zquez-Cend{\'o}n, and Alberto~P
  Mu{\~n}uzuri.
\newblock Interactions between hosts affect virus competition mechanism within
  an infectious strain.
\newblock {\em Chaos, Solitons \& Fractals}, 170:113344, 2023.

\bibitem{kermack1927contribution}
William~Ogilvy Kermack and Anderson~G McKendrick.
\newblock A contribution to the mathematical theory of epidemics.
\newblock {\em Proceedings of the royal society of london. Series A, Containing
  papers of a mathematical and physical character}, 115(772):700--721, 1927.

\bibitem{diekmann2000mathematical}
Odo Diekmann and Johan Andre~Peter Heesterbeek.
\newblock {\em Mathematical epidemiology of infectious diseases: model
  building, analysis and interpretation}, volume~5.
\newblock John Wiley \& Sons, 2000.

\bibitem{hethcote1989three}
Herbert~W Hethcote.
\newblock Three basic epidemiological models.
\newblock In {\em Applied mathematical ecology}, pages 119--144. Springer,
  1989.

\bibitem{ndairou2020mathematical}
Fa{\"\i}{\c{c}}al Nda{\"\i}rou, Iv{\'a}n Area, Juan~J Nieto, and Delfim~FM
  Torres.
\newblock Mathematical modeling of covid-19 transmission dynamics with a case
  study of wuhan.
\newblock {\em Chaos, Solitons \& Fractals}, 135:109846, 2020.

\bibitem{gomes2004infection}
M~Gabriela~M Gomes, Lisa~J White, and Graham~F Medley.
\newblock Infection, reinfection, and vaccination under suboptimal immune
  protection: epidemiological perspectives.
\newblock {\em Journal of theoretical biology}, 228(4):539--549, 2004.

\bibitem{katriel2010epidemics}
Guy Katriel.
\newblock Epidemics with partial immunity to reinfection.
\newblock {\em Mathematical biosciences}, 228(2):153--159, 2010.

\bibitem{wangersky1978lotka}
Peter~J Wangersky.
\newblock Lotka-volterra population models.
\newblock {\em Annual Review of Ecology and Systematics}, 9:189--218, 1978.

\bibitem{fabre2012modelling}
Fr{\'e}d{\'e}ric Fabre, Josselin Montarry, J{\'e}r{\^o}me Coville, Rachid
  Senoussi, Vincent Simon, and Benoit Moury.
\newblock Modelling the evolutionary dynamics of viruses within their hosts: a
  case study using high-throughput sequencing.
\newblock {\em PLoS Pathogens}, 8(4):e1002654, 2012.

\bibitem{barrat2008dynamical}
Alain Barrat, Marc Barthelemy, and Alessandro Vespignani.
\newblock {\em Dynamical processes on complex networks}.
\newblock Cambridge university press, 2008.

\bibitem{estrada2012structure}
Ernesto Estrada.
\newblock {\em The structure of complex networks: theory and applications}.
\newblock American Chemical Society, 2012.

\bibitem{mata2020complex}
Ang{\'e}lica Sousa~da Mata.
\newblock Complex networks: a mini-review.
\newblock {\em Brazilian Journal of Physics}, 50:658--672, 2020.

\bibitem{van2010graph}
Maarten Van~Steen.
\newblock Graph theory and complex networks.
\newblock {\em An introduction}, 144(1), 2010.

\bibitem{bansal2010dynamic}
Shweta Bansal, Jonathan Read, Babak Pourbohloul, and Lauren~Ancel Meyers.
\newblock The dynamic nature of contact networks in infectious disease
  epidemiology.
\newblock {\em Journal of biological dynamics}, 4(5):478--489, 2010.

\end{thebibliography}

\end{document}